\documentclass{osa-article}
\usepackage{physics}
\usepackage{amsmath}
\journal{osac}

\articletype{Research Article}
\begin{document}

\title{Generation of Highly Pure Single-Photon State at Telecommunication Wavelength}

\author{Akito Kawasaki,\authormark{1} Kan Takase,\authormark{1} Takefumi Nomura,\authormark{1} Sigehito Miki,\authormark{2,3} Hirotaka Terai,\authormark{2} Masahiro Yabuno,\authormark{2} Fumihiro China,\authormark{2} Warit Asavanant,\authormark{1} Mamoru Endo,\authormark{1} Jun-ichi Yoshikawa\authormark{1,4} and Akira Furusawa\authormark{1,4,*}}

\address{\authormark{1}Department of Applied Physics, School of Engineering, The University of Tokyo, 7-3-1 Hongo, Bunkyo-ku, Tokyo 113-8656, Japan\\
\authormark{2}Advanced ICT Research Institute, National Institute of Information and Communications Technology,588-2 Iwaoka, Nishi-ku, Kobe, Hyogo 651-2492, Japan\\
\authormark{3}Graduate School of Engineering, Kobe University, 1-1 Rokkodai-cho, Nada-ku, Kobe, Hyogo 657-0013, Japan\\
\authormark{4}Optical Quantum Computing Research Team, RIKEN center for Quantum Computing, 2-1 Hirosawa, Wako, Saitama 351-0198, Japan}

\email{\authormark{*}akiraf@ap.t.u-tokyo.ac.jp} %% email address is required

% \homepage{http:...} %% author's URL, if desired

%%%%%%%%%%%%%%%%%%% abstract %%%%%%%%%%%%%%%%
%% [use \begin{abstract*}...\end{abstract*} if exempt from copyright]

\begin{abstract}
Telecommunication wavelength with well-developed optical communication technologies and low losses in the waveguide are advantageous for quantum applications.
However, an experimental generation of non-classical states called non-Gaussian states at the telecommunication wavelength is still underdeveloped.
Here, we generate highly-pure-single-photon states, one of the most primitive non-Gaussian states, by using a heralding scheme with an optical parametric oscillator and a superconducting nano-strip photon detector.
The Wigner negativity, the indicator of non-classicality, of the generated single photon state is -0.228 ± 0.004, corresponded to $85.1\pm0.7\%$ of single photon and the best record of the minimum value at all wavelengths. 
The quantum-optics-technology we establish can be easily applied to the generation of various types of quantum states, opening up the possibility of continuous-variable-quantum-information processing at the telecommunication wavelength.
\end{abstract}

%%%%%%%%%%%%%%%%%%%%%%%%%%  body  %%%%%%%%%%%%%%%%%%%%%%%%%%
\section{Introduction}
A continuous variable quantum information processing (CV QIP) with optical wave packets \cite{RevModPhys.77.513,takeda2019toward} is one of the most promising approaches to realize a fault tolerant universal quantum computation.
It is known that universal CV QIP requires highly non-classical quantum states with negative values of the Wigner function called Wigner negativity \cite{bib20,PhysRevLett.97.110501}.
Although Quantum states with non-Gaussian Wigner functions are defined as non-Gaussian states\cite{PRXQuantum.2.030204}, in the context of CV QIP, those with Wigner negativities are typically called non-Gaussian states.
Wigner negativity is sensitive to optical losses and deteriorate rapidly with increasing optical losses.
Therefore, the accurate generation and observation of quantum states with Wigner negativity is an important research issue for the realization of CV QIP.
Experimentally, non-Gaussian states have been generated on various platforms such as quantum dots \cite{doi:10.1063/5.0010193} and cavity quantum electrodynamics \cite{Nogues1999,Delglise2008} and ion traps  \cite{PhysRevLett.77.4281,doi:10.1126/science.272.5265.1131}.
Especially, a methodology called a heralding scheme\cite{doi:10.1063/9780735424074} is suitable for an optical CV QIP since it has a high degree of freedom in terms of the quantum states that can be generated, and can generate quantum states on a traveling wave.
By using this method, the generation of  non-Gaussian states such as Schr\"{o}dinger cat states \cite{PhysRevLett.97.083604,Wakui:07,PhysRevA.82.031802}, photon-number states \cite{Neergaard-Nielsen:07,Cooper:13} and superposition states up to three photons \cite{Yukawa:13} have been reported.
As an example of Wigner negativity, single-photon states and odd cat states (and any quantum states that are superposition of odd number Fock states) have the theoretical minimum Wigner negativity of $-1/\pi$ ($\approx-0.32$) \cite{doi:10.1063/9780735424074}.
The value of Wigner negativity is degraded by experimental imperfections, such as optical losses, and the record for the minimum of Wigner negativity experimentally  generated with a heralding scheme so far is -0.22 \cite{PhysRevLett.113.013601}.

The wavelengths of light used in conventional researches of the non-Gaussian-state generation are usually based on atomic transitions, and they are typically around 800\,nm based on Ti:Sapphire lasers or 1064\,nm based on Nd:YAG lasers \cite{PhysRevLett.117.110801}. 
However, when integrating optical quantum computers and constructing quantum communication networks, it is desirable to use the C-band telecommunication wavelength that have low loss in a waveguide and a high compatibility with fiber optics-based technologies.
The development of optical communications has raised the level of classical optical techniques in the telecommunication wavelength, but quantum optical techniques, such as the generation of non-Gaussian states, are still being developed.
High-performance photon detection is one of the essential technologies that needs to be developed to realize an optical CV QIP at telecommunication wavelength.
Photon detection is required for non-Gaussian state generation using heralding scheme, and avalanche photodiodes (APDs) made of silicon have been commonly used in experiments at conventional wavelengths.
On the other hand, the energy of a photon at the telecommunication wavelength is so small that Si-based APDs cannot be used.
So far, wavelength conversion and other efforts have been made to realize photon detection at the telecommunication wavelength using Si-based APDs \cite{PhysRevA.95.061802,Namekata2010}. 
However, due to the conversion in efficiency and the large number of fake clicks caused by pump light, the observed Wigner negativity was down to -0.06 in the case of single-photon-like state \cite{PhysRevA.95.061802}.
This result shows that more improvements are required in the generation of non-Gaussian states at the telecommunication wavelength, comparing to the record of -0.22 achieved at a wavelength of 860\,nm \cite{PhysRevLett.113.013601}.

As another promising photon detector at the telecommunication wavelength, superconducting nano-strip photon detectors (SNSPDs) have been developed\cite{doi:10.1063/1.1388868,Natarajan_2012,Marsili2013}.
However, SNSPDs have not been common in the experiment of an optical CV QIP.
In the previous research, we have succeeded in generating quantum states with Wigner negativity in a system at only  telecommunication wavelength (1545\,nm) without a wavelength conversion for the first time by using a SNSPD \cite{Takase:22} and a waveguide optical parametric amplifier (OPA).
The minimum value of Wiger negativity of the generated quantum state in this experiment is recorded to be -0.12.
Meanwhile, waveguide OPAs have the challenges of large optical losses, meaning that more refinements must be made before they can be applied to high-purity generation of non-Gaussian states.
Another widely-used squeezed light sources, optical parametric oscillators (OPOs), have much lower losses than waveguide OPAs.
Moreover, OPOs can also be used for various applications such as an all-optical quantum memory \cite{PhysRevX.3.041028,Yoshikawamemory} and a real-time measurement \cite{PhysRevLett.116.233602,PhysRevApplied.15.024024}, making them indispensable devices for an optical CV QIP.

In this paper, we construct an extremely low-loss experimental system at the wavelength of 1545\,nm by combining an OPO and a SNSPD, and generate single-photon states using a heralding scheme as a demonstration of non-Gaussian-state generation.
In particular, we use a type of OPO called asymmetric optical parametric oscillator (AOPO) \cite{PhysRevApplied.15.024024}, which is useful for extracting two-mode squeezed states (TMSSs) generated in the orthogonal polarization modes.
We verify the generated single-photon state by reconstructing the Wigner function using a homodyne measurement. Wigner function is sensitive to optical losses and enables a complete characterization of quantum state in phase space, making it suitable for an optical CV QIP. This is in contrast to the second order coherence $g^{(2)}$ which, although being widely used in the evaluation of the single-photon state, is unaffected by optical losses and is unsuitable for an optical CV QIP \cite{scully_zubairy_1997}.
The Wigner negativity of the generated single-photon state is -0.228 ± 0.004. 
To the best of our knowledge, it is the world record of the minimum Wigner negativity experimentally observed without any loss corrections at all wavelengths.
This is the first demonstration of highly pure non-Gaussian state generation using OPO at the telecommunication wavelength, and is an important milestone toward the realization of an optical CV QIP at the telecommunication wavelength.

\section{Experimental details}
In this section, we first give an overview of single-photon states and how they are generated by a heralding scheme, followed by a description of the experimental setup.
\subsection{Single-photon states with optical losses}
Arbitrary quantum state of light can be completely described by a Wigner function.
The Wigner function represents the quasi-probability distribution on the phase space spanned by two canonically conjugate operator $\hat{x},\hat{p}$ called quadrature amplitudes.
Here, the quadrature amplitudes $\hat{x},\hat{p}$ satisfy the commutation relation $\lbrack\hat{x},\hat{p}\rbrack=i$ when $\hbar=1$, and the annihilation operator $\hat{a}$ is given by $\hat{a}=(\hat{x}+i\hat{p})/\sqrt{2}$.
The Wigner function of single-photon states have a negative value of $-1/\pi$ at the origin, and this negative value is an evidence of the non-classicality of the single-photon state.
The density operator of the single-photon states with the optical loss $L$ is expressed as a mixture with the vacuum state and given by
\begin{equation} 
\label{LossSPeq}
\hat{\rho}_{\rm LSP}=L\dyad{0}{0}+(1-L)\dyad{1}{1}.
\end{equation} 
Since the Wigner function of the vacuum state takes the value of $1/\pi$ at the origin, the Wigner negativity at the origin of the mixed state is expressed as $(1-2L)/\pi$.
Equation \ref{LossSPeq} shows that the Wigner negativity of the generated quantum state decreases monotonically with optical losses, and that the Wigner negativity is completely lost at $L>0.5$.
Therefore, it is important to construct a low-loss experimental system for the generation of single-photon states with high purity and large Wigner negativity.

\subsection{Heralding generation of single-photon states}
Single-photon states can be generated by a heralding scheme using photon correlations of TMSSs.
TMSS ($\ket{\psi}_{\rm TMSS}$) in the weak pump limit is given by
\begin{equation}
\ket{\psi}_{\rm TMSS}\sim\ket{0}_1\ket{0}_2+\lambda\ket{1}_1\ket{1}_2,
\end{equation}
where $\lambda$ is a constant that depends on the pump power.
When a photon detection is performed on mode 2, the quantum states generated in mode 1 become
\begin{equation}
 _2\bra{0}\hat{a}_2\ket{\psi}_{\rm TMSS}\rightarrow\lambda\ket{1}_1.
\end{equation}
In other words, photon detection on mode 2 "heralds" single-photon generation in mode 1.

In the heralding scheme using continuous wave (CW) light, generated quantum states are defined in temporal modes with a specific shape.
The temporal mode of the generated quantum state can be derived by the following calculation\cite{doi:10.1063/9780735424074}.
First, TMSS is given by using a photon-pair annihilation operator $\hat{P}_{12}$ as follows.
\begin{equation}
\hat{S}_{12}\ket{0}_1\ket{0}_2=\exp\left(\hat{P}_{12}-\hat{P}_{12}^{\dag}\right)\ket{0}_1\ket{0}_2,
\label{2modesqeq}
\end{equation}
\begin{equation}
\hat{P}_{12}^{\dag}=\frac{1}{\sqrt{2}}\iint dt dt^{\prime}r_{12}(t-t^{\prime})\hat{A}_1^{\dag}(t)\hat{A}_2^{\dag}(t^{\prime}),
\end{equation}
where $r_{12}(t)$ is the time correlation function of photon pairs.
In the weak-pump limit, Eq. \eqref{2modesqeq} can be replaced as follows
\begin{equation}
\hat{S}_{12}\ket{0}_1\ket{0}_2\approx\ket{0}_1\ket{0}_2-\iint dt dt^{\prime}r_{12}(t-t^{\prime})\hat{A}_1^{\dag}(t)\hat{A}_2^{\dag}(t^{\prime})\ket{0}_1\ket{0}_2.
\label{weakpumpeq}
\end{equation}
Considering that the two modes are spatially separated and a photon detection is performed at time $t_0$ on mode 2 which has a filter with an impulse response $g(t)$, the heralded quantum states $\ket{\psi}_1$ of mode 1 is given by
\begin{equation}
\ket{\psi}_1=\int d t\left(\int d \tau {r}_{12}\left(t-t_{0}-\tau\right) g(-\tau)\right) \hat{A}_{1}^{\dagger}(t)\ket{0}_1.
\end{equation}
This result shows that photon detection on one side of the TMSS can herald single-photon states with a temporal mode $h(t)$ on the other side, where $h(t)$ is given by
\begin{equation}
h(t)=\mathcal{N}\lbrack{r}_{12}*g^r(t-t_0)\rbrack.
\end{equation}
Note that $\mathcal{N}\lbrack\cdot\rbrack$ is a normalizing operation and $g^r(t)$ is defined by $g^r(t)=g(-t)$. $*$ represents a convolution.

\subsection{Experimental setup and analysis}

\begin{figure}[h!]
\centering
\includegraphics[width=14cm]{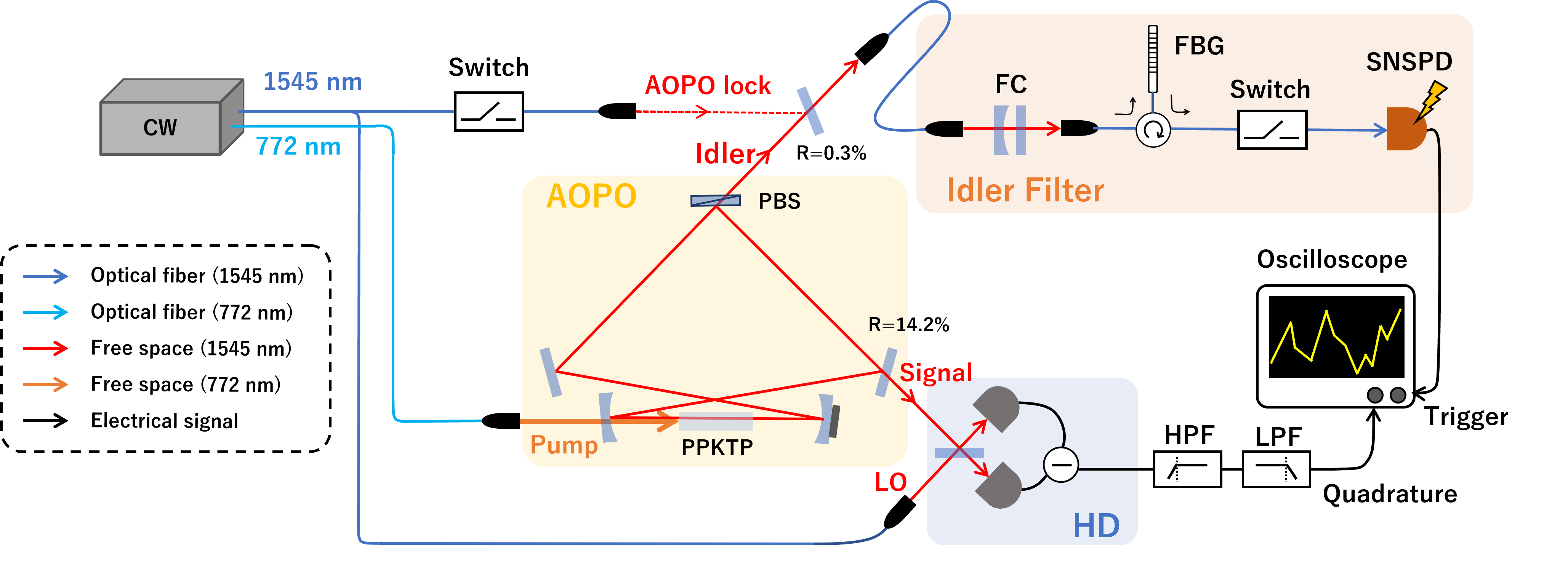}
\caption{Experimental setup of single-photon generation at the telecommunication wavelength of 1545 nm. CW, Continuous wave; AOPO, Asymmetric Optical Parametric Oscillator; FC, Filter Cavity; FBG, Fiber Bragg Grating; SNSPD, Superconducting Nano-Strip Photon Detector; HD, Homodyne Detector; LO, Local Oscillator; HPF, High Pass Filter; LPF, Low Pass Filter.}
\label{setuppic}
\end{figure}
Figure \ref{setuppic} shows the setup of the experiment. 
In this experiment, single-photon states are generated by a heralding scheme.
TMSS is generated in orthogonal polarization modes (signal mode and idler mode) using OPO with a nonlinear optical crystal of type\,$\rm{I\hspace{-.01em}I}$.
The single-photon states are heralded in the signal mode by photon detection using a SNSPD in the idler mode. 
The fundamental laser is a CW light at the wavelength of 1545\,nm.

The squeezed light source is an AOPO, in which one of the mirrors in the OPO is a polarization beam splitter (PBS), so that s-polarized light resonates in the AOPO while p-polarized light is immediately extracted from the PBS.
Since TMSS is generated in orthogonal polarization components, AOPO can be used to spatially separate and extract signal (s-polarization) and idler (p-polarization) light. 
The Free spectral range (FSR) and half width half maximum (HWHM) of the AOPO used in this experiment are 300\,MHz and 3.7\,MHz, respectively.
The nonlinear optical crystal is periodically paled $\rm{KTiOPO_4}$ (PPKTP), whose crystal length is 20\,mm and the phase matching bandwidth is approximately 150\,GHz.
The pump light is CW light at 772.5\,nm generated by second harmonic generation (not shown in Fig. \ref{setuppic}) of the fundamental laser.
The transmittance of the output coupler is 14.2\%, and the round-trip loss inside the AOPO including the crystals is 0.22\%.
The cavity length of the AOPO is locked by lock light injected from the reverse path of the signal and idler light.

Frequency filters are applied to the ider mode in order to utilize only resonance peaks component around the carrier frequency.
The idler filter consists of a filter cavity and a fiber Brag grating (FBG).
The FSR and HWHM of the filter cavity are 8.5\,GHz and 8.2\,MHz, respectively and the HWHM of the FBG is 3.6\,GHz.

The SNSPD used in the experiment is made of NbTiN \cite{Miki:17}. The SNSPD is installed into an adiabatic demagnetization refrigerator and operated at the temperature of 3\,K.
Dark counts of the SNSPD is about 30 counts per a second (cps).
The timing jitter is about 100\,ps, which is sufficiently small compared to the wave-packet width of the single-photon states generated in this experiment.

When classical light is mixed in, the information of weak quantum light is destroyed or the SNSPD is saturated, making it difficult to conduct accurate experiments.
Therefore, in this experiment, the control phase, in which the AOPO and the filter cavity are locked using classical light, and the measurement phase, in which data acquisition is performed, are separated in a time domain. 
One cycle of the measurement is $640\,\rm{\mu s}$, and the measurement phase is $180\,\rm{\mu s}$ per one cycle.
The control and measurement phases are switched by placing  optical switch made of an Acousto-Optic Modulators (AOMs) at the output of the lock light and just before the SNSPD.

\begin{figure}[h!]
\centering
\includegraphics[width=7cm]{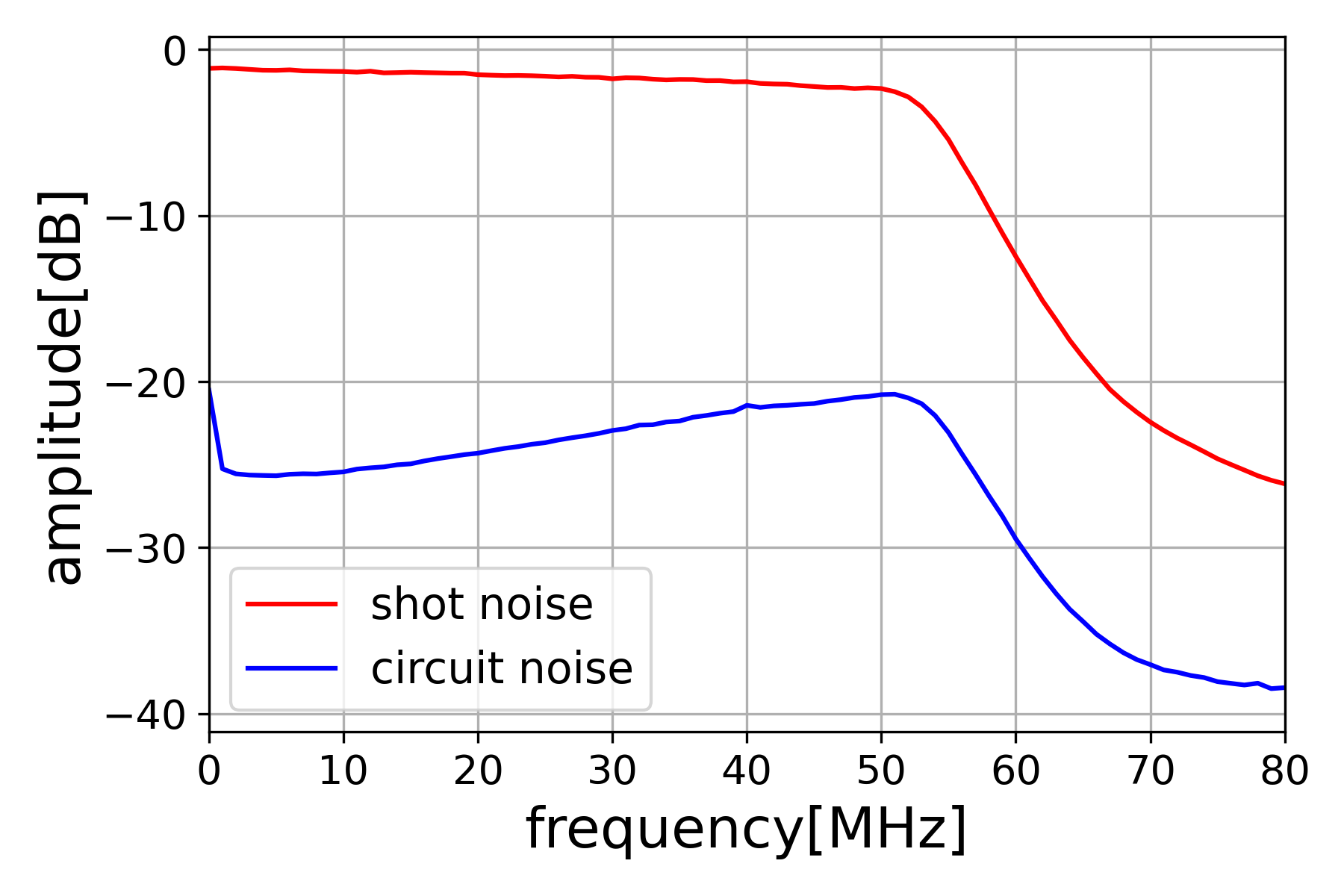}
\caption{The spectrum of shot noise level and circuit noise of homodyne detector when the LO power is set to 5 mW. The components above 50 MHz are cut off by the LPF.}
\label{clearancepic}
\end{figure}

The generated quantum states are evaluated by a homodyne measurements. 
The bandwidth of the homodyne detector is approximately 200 MHz. 
The photodiode of the homodyne detector is made of InGaAs and the quantum efficiency is about 97\%.
The power of the Local oscillator (LO) is set to 5\,mW. Figure \ref{clearancepic} shows the noise lebel of the shot noise and circuit noise when LO power is 5 mW.
The clearance between the shot noise level and the circuit noise is more than 20\,dB. 
The output of the homodyne detector is filtered with a high pass filter (HPF) whose cutoff frequency is 10\,kHz to remove noise around career frequency and a low pass filter (LPF) whose cutoff frequency is 50\,MHz to remove components at sufficiently higher frequency than the quantum states.

The analysis procedure is as follows. 
In this experiment, 20,000 data frames are acquired with randomizing the phase of the homodyne measurement.
The temporal mode of the quantum state is estimated by performing a principal component analysis (PCA) \cite{abdi2010principal} on the experimental data \cite{PhysRevLett.111.213602,PhysRevLett.109.033601,PhysRevA.99.033832}. 
The quadrature amplitude $\hat{x}_h$ defined in temporal mode $h(t)$ is given by
\begin{equation}
    \hat{x}_h=\int h(t)\hat{x}(t)dt,
\end{equation}
where $\hat{x}(t)$ represents the quadrature amplitude at each time of $t$.
By calculating the values of the quadrature amplitudes in each frame, the quantum state can be reconstructed by the quantum tomography \cite{RevModPhys.81.299}.
We assume that the quantum states in this experiment are phase insensitive when implementing the quantum tomography to reconstruct the quantum states.

\section{Experimental Results}
In this experiment, the power of the pump light was set to 2\,W. At this time, the total counts of SNSPD is 3,000\,cps, of which the dark counts is 30\,cps and the stray counts caused by lock light, pump light, or stray light is 60\,cps.

\begin{figure}[h!]
\centering
\includegraphics[width=14cm]{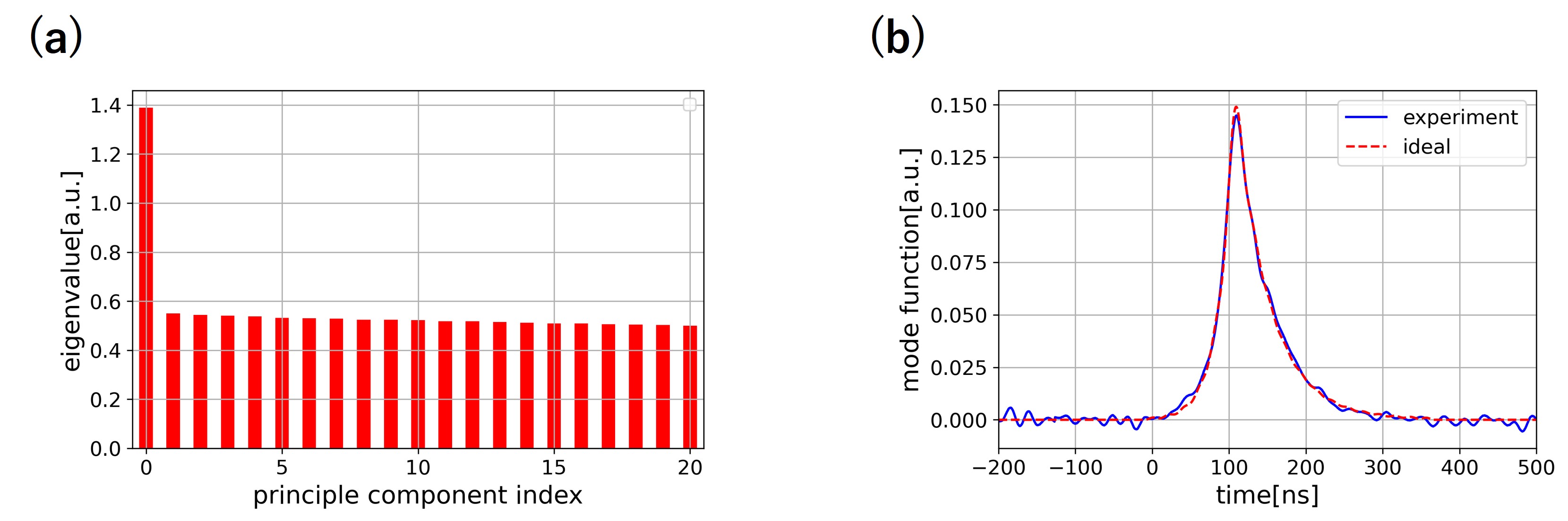}
\caption{The results of PCA. (a)Eigenvalues of each component. (b)Estimated temporal mode function(the principle component obtained from PCA).}
\label{modefunctionpic}
\end{figure}

Figure \ref{modefunctionpic} shows the eigenvalues and the first eigenfunction of PCA.
Figure \ref{modefunctionpic} (a) indicates the eigenvalue of the principal components are more dominant than the other components.
The experimentally estimated temporal mode function, the first eigenfunction of PCA, shows a good agreement with the theoretical curve, which takes into account the AOPO bandwidth, idler filters, and electrical frequency filters in the homodyne measurement.
A mode match between theoretical line and experimental line is 99.5\%.

\begin{figure}[h!]
\centering
\includegraphics[width=7cm]{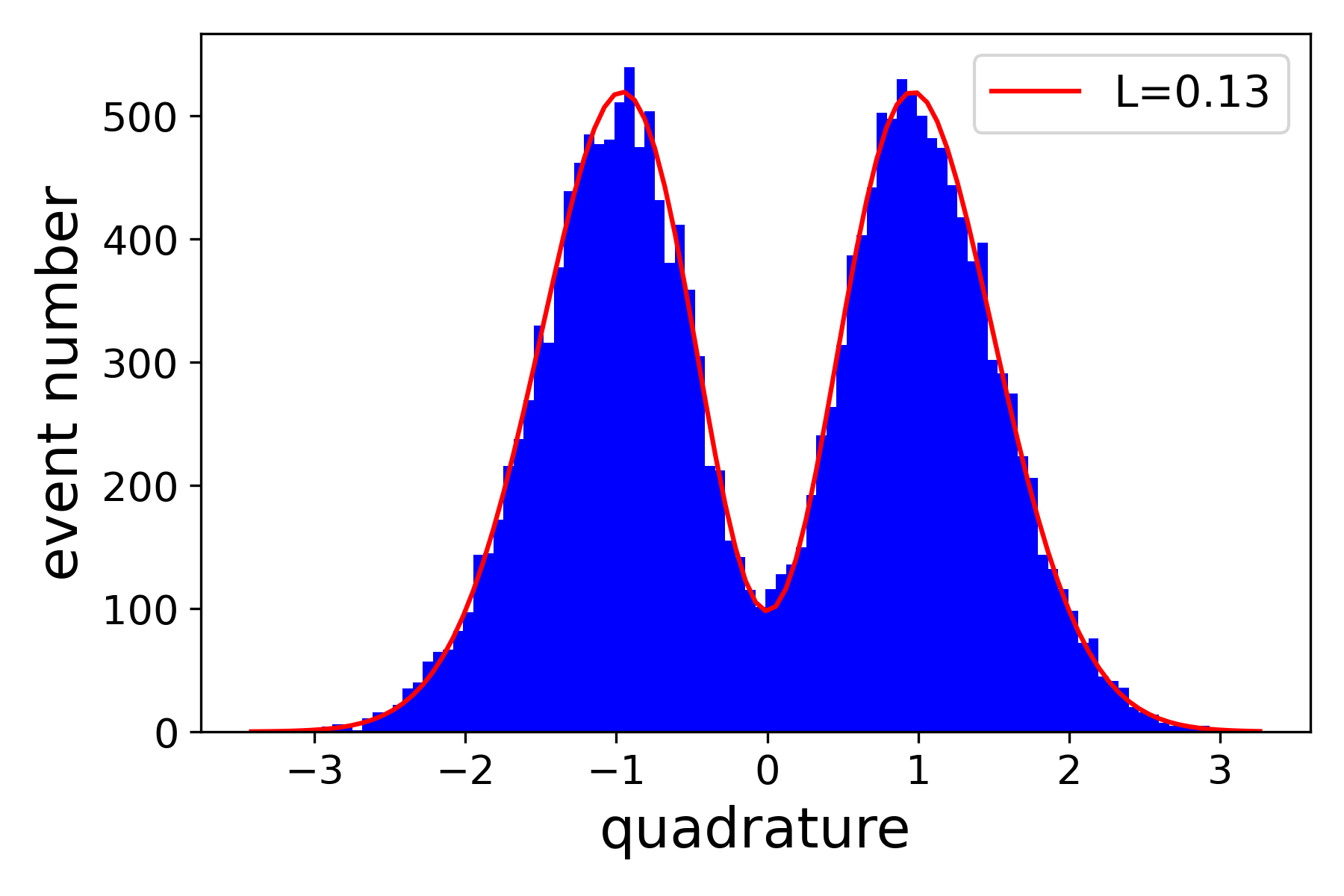}
\caption{Histogram of experimentally obtained quadrature amplitudes. The red line is the theoretical line of the marginal distribution function of quadrature amplitudes of the single-photon states when the optical loss $L$ is 13\%.}
\label{pp_hist_with_losspic}
\end{figure}

The histogram of the estimated quadrature amplitude in the estimated temporal mode is shown in Fig. \ref{pp_hist_with_losspic}.
The histogram shows the dip around the origin, which is a characteristic of single-photon states.
In the case of the ideal single-photon state, the origin value of a quadrature distribution should be zero, but the experimental results are not zero due to the optical losses and they are in good agreement with the theoretical marginal distribution function for the case of $L=0.13$.

\begin{figure}[h!]
\centering
\includegraphics[width=14cm]{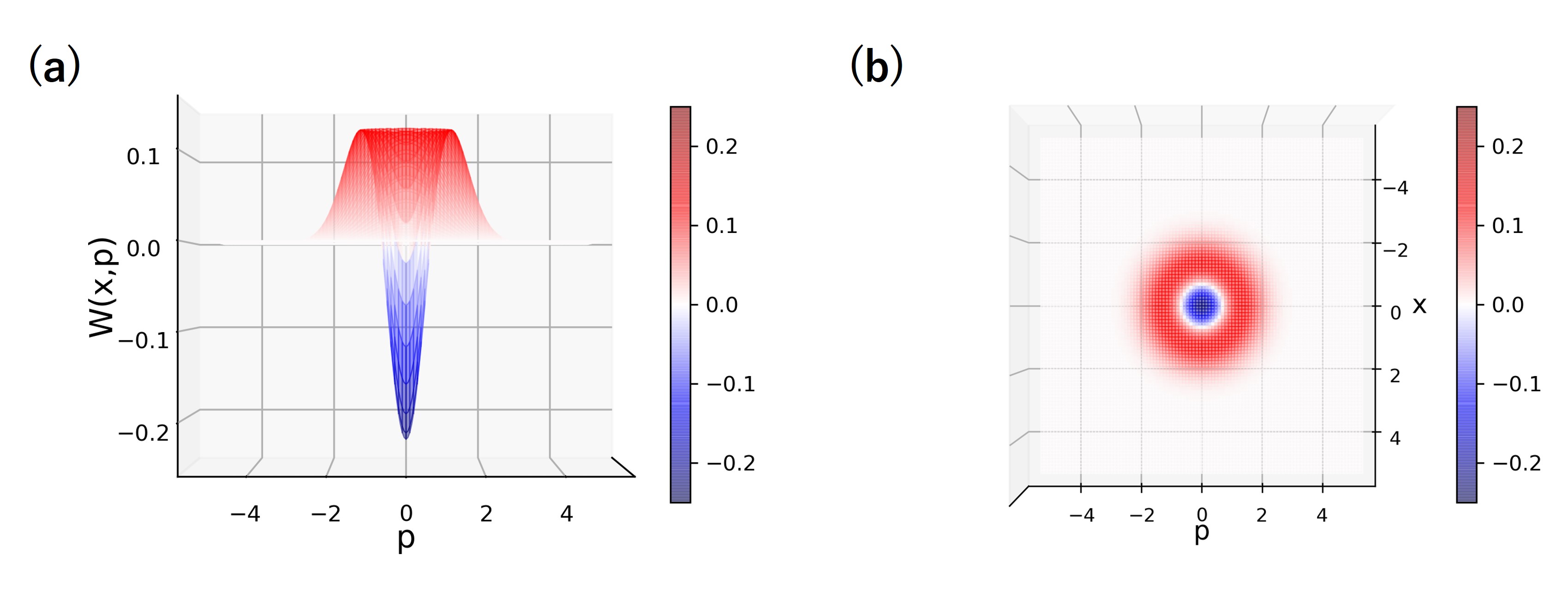}
\caption{The Wigner functions of single photon states generated by a heralding scheme. The value of the origin is -0.228±0.004. (a) side view, (b)top view}
\label{Wignerpic}
\end{figure}

The Wigner function of the quantum state reconstructed by quantum tomography is shown in Fig. \ref{Wignerpic}.
The negative value at the origin is -0.228±0.004. 
The bootstrap method \cite{10.1214/aos/1176344552} is used to estimate the error of the quantum tomography.

\begin{figure}[h!]
\centering
\includegraphics[width=7cm]{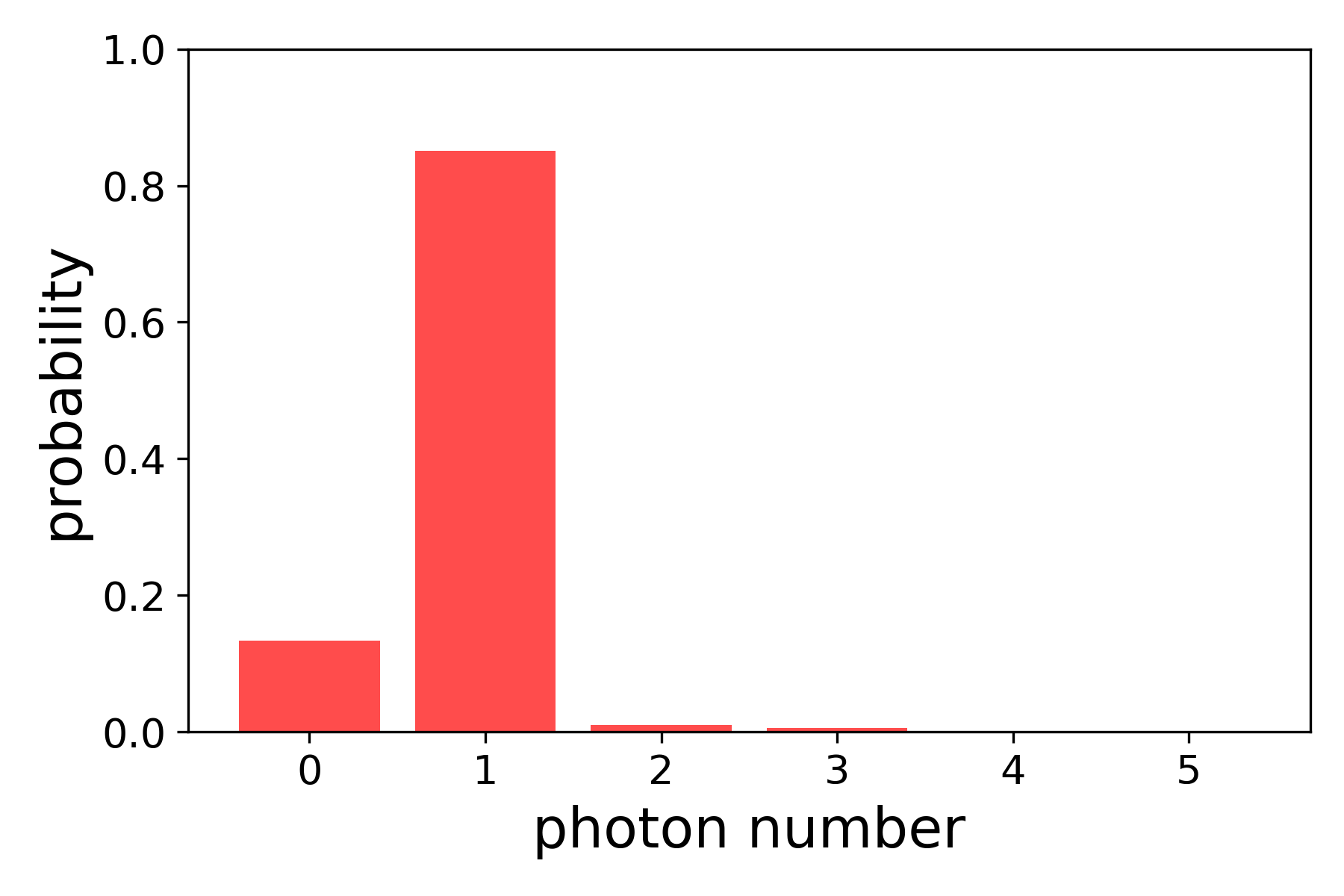}
\caption{Breakdown of the contribution of the Fock state obtained by quantum tomography. The largest contribution is single photon ($85.1\pm0.7\%$).}
\label{photon_numberpic}
\end{figure}

Figure \ref{photon_numberpic} shows the distribution of photon numbers of quantum states obtained from quantum tomography.
The proportion of the n-photon component $p_n$ is $p_0=13.3\pm0.5\%$, $p_1=85.1\pm0.7\%$, $p_2=1.0\pm0.7\%$, $p_3=0.5\pm0.4\%$.
The two or more photon components are thought to originate from higher order terms in Eq.\eqref{weakpumpeq} that arises when the pump is strong and the weak pump approximation is not valid.

Finally, a breakdown of the optical losses estimated from the experimental system is summarized in table \ref{losstable}.
The sum of these estimated optical losses is 13\%, indicating that the experimental result of 13.3\% for the vacuum component is a reasonable value.
\begin{table}
 \centering
  \begin{tabular}{lc}
   \hline 
   dark count (loss equivalent)& $1\%$\\
   stray count (loss equivalent)& $2\%$\\
   optical loss of AOPO& $2\%$\\
   propagation loss& $3\%$\\
   mode mismatch of HD & $2\%$\\
   inefficiency of photodiodes of HD& $3\%$\\
   circuit noise of HD& $1\%$\\
   \hline
   total& $13\%$\\
   \hline
  \end{tabular}
  \caption{Estimated loss budget.}
  \label{losstable}
\end{table}

\section{Conclusion}
In this research, we have succeeded in generating highly pure single-photon states at the telecommunication wavelength. 
The proportion of the single-photon state in the experimentally obtained quantum states is $85.1\pm0.7\%$, and the Wigner negativity is -0.228±0.004 which is the best record of minimum Wigner negativity at the telecommunication wavelength. 
Furthermore, to the best of our knowledge, this value also breaks the previous record of Wigner negativity -0.22 \cite{PhysRevLett.113.013601} in a single-photon-state generation at all wavelengths.
This result is achieved by constructing an unprecedentedly low-loss experimental system of the heralding scheme.
In our previous experiments using a waveguide OPA and a SNSPD \cite{Takase:22}, the optical losses of the squeezed light source was about 10\%, while in this experiment, we succeeded in reducing the optical losses of AOPO to 2\% by a low-loss coatings of mirrors and a crystal.

As a future outlook, higher purity single-photon states can be generated by improving some parts of the experimental system.
The main cause of the dark counts in the SNSPD used in this experiment is estimated to be photons around the wavelength of $2\,\mu \rm{m}$ due to spontaneous emissions in the optical fiber.
The dark counts could be improved by introducing AR coatings or bandpass filters at SNSPD.
Using a free-space coupled SNSPD is also one of the most promising ways to reduce dark counts\cite{Mueller:21}.
In addition, further purification could be possible by improving the detection efficiency of SNSPDs and the quantum efficiency of photodiodes in homodyne detectors.
In particular, SNSPDs with detection efficiencies of more than $99\%$ at the telecommunication wavelength already been developed\cite{doi:10.1063/5.0039772}.

The combination of OPO and SNSPD is applicable not only to single-photon state generation but also to the heralding generation of various quantum states. 
Especially, the importance of low-loss systems becomes even more critical in the case of complex quantum state generation where photon-number-resolving detection is required, for example Gottesman-Kitaev-Preskill state \cite{PhysRevA.64.012310} or Schr\"{o}dinger cat states with large amplitude \cite{PhysRevA.103.013710}.
Based on this research, it is expected that more complex quantum states useful for an optical CV QIP can be generated with high purity.

In this research, we have demonstrated a new quantum optics technology that realizes highly pure non-Gaussian state generation toward CV QIP at the telecommunication wavelength.
This research is a technological innovation that links classical optics technology, developed along with optical communications, and quantum computation, opening up the possibility of an optical CV QIP at the telecommunication wavelength.

\begin{backmatter}
\bmsection{Fundings}
Japan Society for the Promotion of Science KAKENHI (18H05207, 18H01149, 20J10844, 20K15187); Japan Science and Technology Agency (JPMJMS2064,JPMJMS2066)

\bmsection{Acknowledgments}
The authors acknowledge supports form UTokyo Foundation and donations from Nichia Corporation of Japan. W.A. and M.E. acknowledge supports from Research Foundation for Opto-Science and Technology. A.K. acknowledges financial support from The Forefront Physics and Mathematics Program to Drive Transformation (FoPM).

\bmsection{Disclosures}
The authors declare no conflicts of interest.

\bmsection{Data availability}
 Data underlying the results presented in this paper are not publicly available at this time but may be obtained from the authors upon reasonable request.

\end{backmatter}

%%%%%%%%%%%%%%%%%%%%%%% References %%%%%%%%%%%%%%%%%%%%%%%%%
\bibliography{ref}
\end{document}